\documentclass[11pt,twoside]{article}
\usepackage{asp2010}

\resetcounters

\begin{document}

\title{Observational consequences of the Partially Screened Gap}
\author{Andrzej~Szary, George~Melikidze, and Janusz~Gil
\affil{Kepler Institute of Astronomy, University of Zielona G\'{o}ra, Lubuska
    2, 65-265, Zielona G\'{o}ra, Poland}
}

\begin{abstract}
    Observations of the thermal X-ray emission from old radio pulsars implicate
    that the size of hot spots is much smaller then the size of  the polar cap
    that follows from the purely dipolar geometry of pulsar magnetic field.
    Plausible explanation of this phenomena is an assumption that the magnetic
    field at the stellar surface differs essentially from the purely dipolar field.
    Using the conservation of the magnetic flux through the area bounded by open
    magnetic field lines we can estimate the surface magnetic field as of the 
    order of $10^{14}$G. Based on observations that the hot spot temperature is 
    about a few million Kelvins the Partially Screened Gap (PSG) model was 
    proposed which assumes that the temperature of the actual polar cap equals to 
    the so called critical temperature. We discuss correlation between the 
    temperature and corresponding area of the thermal X-ray emission for a number
    of pulsars.

    We have found that depending on the conditions in a polar cap
    region the gap breakdown can be caused either by the Curvature Radiation (CR)
    or by the Inverse Compton Scattering (ICS). When the gap is dominated by ICS
    the density of secondary plasma with Lorentz factors $10^{2}-10^{3}$  is at 
    least an order of magnitude higher then in a CR scenario. We believe that
    two different gap breakdown scenarios can explain the mode-changing
    phenomenon and in particular the pulse nulling.
    Measurements of the characteristic spacing between sub-pulses ($P_{2}$) and
    the period at which a pattern of pulses crosses the pulse window ($P_{3}$) 
    allowed us to determine more strict conditions for avalanche pair production
    in the PSG. 
    
\end{abstract}

\section{Introduction}

    The Standard model of radio pulsars assumes that there exists the Inner
    Acceleration Region (IAR) above the polar cap where the electric field has
    a component along the opened magnetic field lines. In this region particles
    (electrons and positrons) are accelerated in both directions: outward and
    toward the stellar surface (\cite{1975_Ruderman}). Consequently, outflowing
    particles are responsible for generation of the magnetospheric emission
    (radio and high-frequency) while the backflowing particles heat the
    surface and provide required energy for the thermal emission. The Vacuum
    Gap model assumes that ions cannot be extracted from stellar surface due to
    huge surface magnetic field of a pulsar. On the other hand it predicts
    the surface temperature of few million Kelvins (heating by backflowing
    particles). As shown by \cite{2007_Medin} for such high temperatures the
    ions extraction from surface cannot be ignored. In fact for surface
    temperature few million Kelvins the gap can form only if surface magnetic
    field is much stronger than the dipolar component ($B_s=10^{14} G$).

    The analysis of X-ray radiation is an excellent method to get insight into
    the most intriguing region of the neutron star (NS).
    X-ray emission seems to be a quite common feature of radio pulsars.
    In general X-ray radiation from an isolated NS can consist of
    two distinguishable components: the thermal emission and the 
    nonthermal emission. The thermal emission can originate either from
    the entire surface of cooling NS or the spots around
    the magnetic poles on stellar surface (polar caps and adjacent areas).
    The nonthermal component is usually attributed to radiation produced
    by Synchrotron Radiation and/or Inverse Compton Scattering from
    charged relativistic particles accelerated in the pulsar magnetosphere.
    For most observations it is very difficult to distinguish contribution of
    different components (thermal and nonthermal). To get an information about
    polar cap of radio pulsars we analysed X-ray radiation from old pulsars as
    their surface is already cooled down and their magnetospheric radiation
    (nonthermal component) is also significantly weaker.

    The blackbody fit allows us to obtain directly the temperature ($T_s$)
    of the hot spot. Using the distance ($D$) to the pulsar and the
    luminosity of thermal emission ($L_{bol}$) we can estimate the 
    area ($A_{pc}$) of the hot spot. In most cases  ($A_{pc}$)
    differs from the conventional polar cap area 
    $A_{dp} \approx 6.2 \times 10^4 P^{-1} \,{\rm m^2}$, where $P$
    is the pulsar period. We use parameter $b = A_{dp} / A_{pc}$ to
    describe the difference between $A_{dp}$ and $A_{pc}$.
    Pulsars for which it is possible to determine polar cap size 
    (old NSs) show that the actual polar cap size is much smaller
    ($b\gg 1$) than the size of conventional polar cap (see Tab.
    \ref{tab:results}).
    
    The surface magnetic field can be estimated by the magnetic flux
    conservation law as $b = A_{dp} / A_{pc} = B_s / B_d$, where
    $B_d = 2.02 \times 10^{12} \left ( P \dot{P}_{-15} \right ) ^{0.5}$, and
    $\dot{P}_{-15} = \dot{P}/10^{-15}$ is the period derivative.
    The X-ray observations suggest that surface magnetic field strength
    at polar cap should be of the order of $10^{14}$ G. On the other hand
    we know from radio observations that magnetic field at altitudes where
    radio emission is generated should be dipolar. To meet both these
    requirements Partially Screened Gap model assumes the existence of
    crust-anchored local magnetic anomalies which affect magnetic field
    only on short distances. According to our model the actual surface 
    temperature equals to the critical value ($T_s \sim T_{crit}$)
    which leads to the formation of Partially Screened Gap.

\section{Partially Screened Gap}
    The PSG model assumes existence of heavy (Fe$^{56}$) ions with density near
    but still below corotational charge density ($\rho_{{\rm GJ}}$), thus the
    actual charge density causes partial screening of the potential drop just
    above the polar cap. The degree of shielding can be described by shielding
    factor $\eta=1-\rho_{i}/\rho_{{\rm GJ}}$,
    where $\rho_{i}$ is the charge density of heavy ions in the gap.
    The thermal ejection of ions from surface causes partial screening of the
    acceleration potential drop $\Delta V=\eta\Delta V_{max}$, where
    $\Delta V_{max}$ is the potential drop in vacuum gap.
    Using calculations of \cite{2007_Medin} we can express the dependence of
    the critical temperature on pulsar parameters as
    $T_{{crit}}=1.1\times10^{6}\left(B_{14}^{1.1}+0.3\right)$, where
    $B_{14}=B_{s}/10^{14}$, $B_{s}=bB_{d}$ is surface magnetic field
    (applicable only if hot spot is observed i.e. $b>1$).

    The actual potential
    drop $\Delta V$ should be thermostatically regulated and there should be
    established a quasi-equilibrium state, in which heating due to
    electron/positron bombardment is balanced by cooling due to thermal
    radiation (see \cite{2003_Gil} for more details). The necessary condition for
    formation of this quasi-equilibrium state is
    \begin{equation}
        \sigma T_{s}^{4}=\eta e\Delta Vcn_{{\rm GJ}},
    \end{equation}
    where $\sigma$ is the Stefan-Boltzmann constant, $e$ - the electron charge,
    $n_{{\rm GJ}}=\rho_{{\rm GJ}}/e= 6.93\times10^{12}B_{14}P^{-1}$ is the
    corotational number density.

    Using the Gauss's law and Faraday's law of induction we can find the
    formula for potential drop in a gap region
    \begin{equation}
         \Delta V/h^{2}+ \Delta V/h_{\perp}^{2} =
         4 \pi \eta B_{s}\cos\left(\alpha\right)/cP
         \label{eq:delta_v}
    \end{equation}
    where $h$ is gap height, $h_{\perp}$ is spark width and $\alpha$ is
    the inclination angle between rotation and magnetic axis.
    We have found that the main parameter that determines the process
    responsible for gamma-ray photon emission in gap region is spark width
    ($h_{\perp}$). For narrower sparks (higher shielding factor) acceleration
    potential drop is lower, which results in smaller Lorentz factors of primary
    particles ($\gamma \sim10^{3}-10^{4}$). In this regime the gap will be
    dominated by ICS. Wider sparks (smaller shielding factor) corresponds to
    higher acceleration ($\gamma \sim10^{5}-10^{6}$) and results in gap
    dominated by CR. In this case the particles will be accelerated to higher
    energies before they would upscatter x-ray photons emitted from the hot
    polar cap. As the determination of spark width is not possible by only
    using X-ray data we decided to use radio observations to put more
    strict constrains on PSG model.

\section{The drift model}
    The existence of IAR in general causes rotation of plasma relative to the
    NS. The power spectrum of radio emission must have a feature due to this
    plasma rotation. This feature is indeed observed and it is called
    drifting sub-pulse phenomenon. Using assumption that the spark width and
    distance between sparks are of the same order, we can define the drifting
    velocity as
    \begin{equation}
         v_{dr}=2 h_{\perp} / \left ( P P_{3} \right)
    \end{equation}
    where $P_3$ is the period at which a pattern of pulses crosses the pulse
    window (in units of pulsar period). In our model drift is caused by lack
    of charge in IAR, then knowing that
    ${\bf v_{\perp}}=c{\bf \Delta E}\times{\bf B} / B^{2}$ we can use calculation of
    circulation of electric field to find the dependence of drift velocity on
    shielding factor
    \begin{equation}
         v_{dr}= 4 \pi \eta h_{\perp}\cos\alpha/P
    \end{equation}
    Finally we can find dependence of shielding factor on observed drift
    parameters
    \begin{equation}
        \eta = 1/2 \pi P_{3} \cos \alpha
        \label{eq:eta}
    \end{equation}
    Knowing that heating luminosity
    $L_{heat}=\eta n_{{\rm GJ}}\left(\Delta Ve\right)c\pi R_{pc}^{2}$ we can
    use Eqs. \ref{eq:delta_v} and \ref{eq:eta} to find the dependence
    of heating efficiency ($\xi=L_{ heat}/L_{sd}$) on sub-pulse drift
    parameters
    \begin{equation}
        \xi \approx 0.15\left(P_{2}^{\circ} / \left ( P_{3} W_{\beta0} \right) \right)^{2},
    \end{equation}
    where $W_{\beta0}$ is the pulse width in degrees calculated with an
    assumption that impact angle is zero ($\beta=0$).
    Thus, radio data allow not only to determine shielding factor (and
    hence width of the sparks, see Eq. \ref{eq:delta_v}) but also observations
    of sub-pulse drift allow to predict polar cap x-ray luminosity.
    Tab. \ref{tab:results} presents observed and derived parameters of PSG for
    pulsars with available radio and x-ray data. Please note that we
    consider only pulsars with visible hot spot component (old NS). Despite the
    fact that sample is very small we still managed to determine that for
    observed pulsars ICS is responsible for gamma-photon generation in IAR.

\begin{table}[ht]
    \caption{Observed and derived parameters of PSG for pulsars with available
    radio observations
    of sub-pulse drift ($P_{2}^{\circ}$, $P_{3}$) and X-ray observations of actual
    polar cap (hot spot). $T_s$, $R_{pc}$ and $B_s$ was chosen to fit $1\sigma$
    uncertainty. Please note that for calculations $\tilde{P}_{2}^{\circ}$ was
    used as the predicted value of sub-pulse separation
    (the observed value is greater than pulse width and can not be
    interpreted as the actual sub-pulse separation). \label{tab:results}}
    \begin{center}
    \begin{tabular}{|l|c|c|c|c|c|c|c|c|c|}
    \hline
         & & &  &  &  &  &  &  & \tabularnewline
        Name  &  $P_{3}$ &  $\eta$ &  $\tilde{P}_{2}^{\circ}$ & $\log \xi$  & $\log \xi_{bol}$  &$T_s$ & $B_s$ & $R_{pc}$  & $h_{\perp}$ \\
             &  {\scriptsize $\left(P\right)$} &   &   {\scriptsize $\left({\rm deg}\right)$}   & {\scriptsize $\left({\rm radio}\right)$}  & {\scriptsize $\left({\rm x-ray}\right)$}  & {\scriptsize $\left ( 10^{6}{\rm K} \right )$} & {\scriptsize $\left ( 10^{14}{\rm G} \right )$} & {\scriptsize $\left({\rm m}\right)$}  & {\scriptsize $\left({\rm m}\right)$}\\
        \hline
        \hline
        B0628--28    & $7.0$  & $0.07$ & $7.6$ & $-4.0$  & $-3.6$  & $2.5$ & $2.0$ & $23$  & $3.9$ \\
        B0834+06   & $2.2$  & $0.15$ & $1.1$ & $-3.6$  & $-3.3$  & $3.0$ & $2.4$ & $20$  & $1.8$ \\
        B0943+10    & $1.8$  & $0.09$ & $8.9$ & $-3.2$  & $-3.3$  & $3.2$ & $2.5$ & $17$  & $2.0$ \\
        B0950+08  & $6.5$  & $0.09$ & $2.8$ & $-5.1$ & $-4.5$  & $2.6$ & $2.1$ & $14$  & $0.7$ \\
        B1133+16  & $3.0$  & $0.09$ & $2.7$ &  $-3.3$ & $-3.1$  & $3.4$ & $2.7$ & $17$  & $2.9$ \\
        B1929+10  & $9.8$  & $0.02$ & $5.2$ & $-5.1$  & $-4.2$  & $4.2$ & $2.0$ & $22$  & $1.6$ \\
        \hline
    \end{tabular}
    \end{center}
\end{table}

    ICS in strong magnetic fields is very efficient process i.e. particle loses
    most of its energy during scattering. This is the cause of very high
    multiplicity, $M$, (number of secondary particles produced by one primary
    particle). The number density of secondary plasma in PSG model can be
    described as $n_{ sec}=\eta n_{{\rm GJ}} M$. ICS dominated gap produces
    two populations of secondary plasma. The first
    population (higher Lorentz factors) is produced when primary particles
    lose most of their energy in ICS process. The second population corresponds
    to particles produced by gamma-ray photons above the gap (lower Lorentz
    factors).


\section{Conclusions}
    To follow both theoretical predictions and observational data PSG model
    was proposed. Recent studies on the model showed that cascade scenario in
    a gap (CR or ICS) strongly depends on spark width. X-ray observations in
    combination with sub-pulse drift analysis allowed to determine that for
    observed pulsars ICS is responsible for gamma-ray photon generation in a
    gap. The exact density of secondary plasma can be
    calculated only by performing full cascade simulation with inclusion of
    heating by backstreaming particles. Nevertheless we can still find
    dependence of multiplicity factor on number of photons upscatterd by one
    primary particle. We were able to find two populations of secondary plasma
    with different energy distribution. It turns out that ICS dominated gap
    creates conditions suitable for generation of radio emission at altitudes
    several tens of stellar radii.


\bibliography{bibliography}

\end{document}